\ifdefined\pdfminorversion
\fi
\documentclass[%
reprint, superscriptaddress,
amsmath,amssymb, aps,
prb,
showkeys, ]{revtex4-2}
\usepackage{graphicx}% Include figure files
\usepackage{dcolumn}% Align table columns on decimal point
\usepackage{bm}% bold math
\usepackage{booktabs}
\usepackage{xcolor}
\colorlet{BLUE}{blue}
\usepackage{threeparttable}

\usepackage[colorlinks=true,allcolors=blue]{hyperref}
\hypersetup{
	pdftitle={Symmetry-Adapted Physical and Vibrational Properties of Ferroelectric Perovskite Oxides: Application to PbZrxTi1-xO3},
	pdfauthor={Sumit Ranjan Maity and Brajesh Tiwari}
}
\makeatletter
\def\@bibdataout@aps{%
	\immediate\write\@bibdataout{%
		@CONTROL{%
			apsrev42Control%
			\longbibliography@sw{%
				,author="48",editor="1",pages="0",title="0",year="1"%
			}{%
				,author="48",editor="1",pages="0",title="",year="1"%
			}%
		}%
	}%
	\if@filesw
	\immediate\write\@auxout{\string\citation{apsrev42Control}}%
	\fi
}%
\makeatother
\begin{document}
	
	\title{Symmetry-Adapted Physical and Vibrational Properties of Ferroelectric Perovskite Oxides: Application to PbZr$_x$Ti$_{1\text{-}x}$O$_3$}
	
	\author{Sumit Ranjan Maity}
	\thanks{Corresponding author: Sumit Ranjan Maity, email: sumit050491@spring8.or.jp}
	
	\affiliation{Japan Synchrotron Radiation Research Institute, Sayo, Hyogo 679-5198, Japan}
	
	\author{Brajesh Tiwari}
	\thanks{Corresponding author: Brajesh Tiwari, email: brajeshtiwari@iitram.ac.in}
	\affiliation{Department of Basic Sciences, Institute of Infrastructure, Technology, Research And Management, Ahmedabad 380026, India.}

	
	\begin{abstract}
		Crystal symmetry governs macroscopic physical properties and lattice dynamics in functional materials. We present a systematic application of tensor analysis and group theory to determine allowed physical-property tensors, vibrational-mode symmetries, and Raman selection rules directly from crystallographic point-group symmetry. The approach is applied to the prototypical ferroelectric PbZr$_x$Ti$_{1-x}$O$_3$ (PZT). Symmetry lowering across the PZT phase diagram increases the number of independent pyroelectric, dielectric, and piezoelectric tensor components and modifies the symmetry classification of Raman-active vibrations. These mode classifications enable symmetry-based decomposition of reported room-temperature powder Raman spectra as a function of composition ($x$) across the morphotropic phase boundary, revealing the tetragonal-to-rhombohedral transition as a continuous redistribution of spectral intensity rather than emergence of new Raman modes. Persistent subpeak structure in selected modes indicates local symmetry breaking due to cation disorder and lattice anharmonicity, underscoring the importance of crystallographic symmetry analysis for interpreting functional and vibrational properties of ferroelectric perovskite oxides.
	\end{abstract}

	
	\maketitle
	
	%\tableofcontents
	
	\section{Introduction}
	
	Crystal symmetry determines the physical properties that a material can exhibit. The existence and anisotropy of ferroelectric responses such as spontaneous polarization, dielectric susceptibility, and piezoelectricity are constrained by crystallographic symmetry, while the same symmetry operations govern lattice vibrations and their spectroscopic selection rules \cite{Nye1985,Dresselhaus2008,Newnham2005}. Consequently, macroscopic property tensors and microscopic vibrational excitations represent complementary manifestations of the same underlying symmetry. Understanding this relationship is particularly important in materials undergoing structural phase transitions, where symmetry breaking simultaneously modifies both functional properties and lattice dynamics.
	
	The connection between symmetry and physical properties is well established through Neumann's principle, which constrains the allowed forms of tensor quantities according to the symmetry of the crystal \cite{Nye1985}. In parallel, group-theoretical methods provide a rigorous framework for classifying lattice vibrations, determining irreducible representations, and establishing Raman and infrared selection rules \cite{BornHuang1954,Herzberg1945,HayesLoudon1978,YuCardona2010}. Modern crystallographic tools, such as the Bilbao Crystallographic Server, further facilitate the systematic determination of symmetry representations, tensor forms, and selection rules \cite{kroumova2003}. Although these approaches are individually well-developed, they are often presented and applied separately, despite originating from the same crystallographic symmetry operations. Treating them together therefore offers a more coherent and transferable route to analyzing symmetry-driven behavior in functional materials.
	
\begin{figure*}[t]
		\centering
		\includegraphics[width=\linewidth]{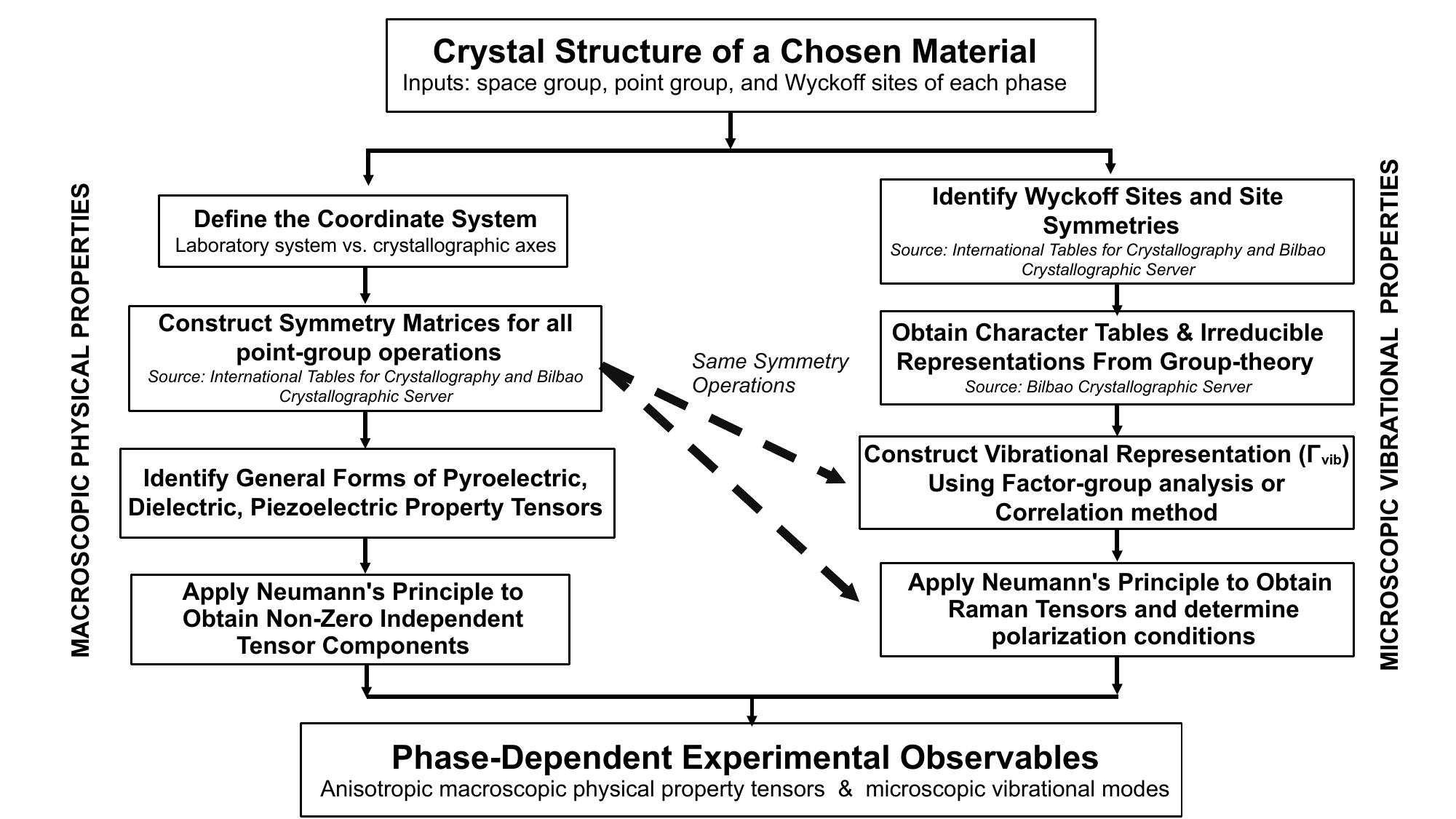}
		\caption{Overview of the symmetry workflow. Starting from the crystal structure of a material, the crystallographic point-group operations are used in two parallel analyses: a macroscopic branch that derives the allowed tensor forms via Neumann’s principle, and a microscopic branch that determines the vibrational modes and Raman tensors. The two branches converge to phase-dependent experimental observables.}
		\label{fig:symmetry_workflow}
		
		\textbf{Alt text:} The symmetry-guided workflow, starting from the crystal structure of a chosen material through two parallel branches that converge to phase-dependent experimental observables.
	\end{figure*}

	In this work, we present a systematic application of established crystallographic methods, organizing them into a parallel workflow as summarized in Fig.~\ref{fig:symmetry_workflow}. Starting from the crystallographic point-group symmetry of a given phase, the workflow divides into two branches. The macroscopic branch applies Neumann's principle to determine the symmetry-allowed forms of the physical-property tensors, while the microscopic branch employs factor-group and correlation methods to classify zone-center vibrational modes, Raman tensors, and selection rules.
	
	PbZr$_x$Ti$_{1-x}$O$_3$ (PZT) provides an ideal system for demonstrating this workflow owing to its rich structural phase diagram and strong coupling between crystal symmetry and functional properties \cite{Jaffe1971,Cohen1992,Noheda2000}. PZT undergoes symmetry-lowering transitions from the cubic paraelectric phase to ferroelectric tetragonal, rhombohedral, and monoclinic phases, each characterized by distinct polarization directions, tensor properties, and vibrational symmetries \cite{Jaffe1971,Cohen1992,Noheda2000}. Of particular interest is the morphotropic phase boundary (MPB), where competing structural instabilities and polarization rotation pathways give rise to enhanced electromechanical responses \cite{Cohen1992,Noheda2000,Schonau2007}. The primary objective is to demonstrate how this rigorous symmetry-based baseline serves as a valuable reference for interpreting complex experimental phenomena, such as local symmetry breaking, phase coexistence, and spectral-weight transfer across the MPB, with PZT serving as a representative system.
	
	Using this framework, we derive the symmetry-allowed pyroelectric, dielectric, and piezoelectric tensor forms for the major structural phases of PZT and determine their corresponding Raman-active vibrational modes, Raman tensors, and selection rules. These symmetry-derived results provide the basis for the analysis of previously reported room-temperature powder Raman spectra across the MPB \cite{maity2023}. Although PZT possesses the simple ABO$_3$ perovskite structure, its Raman spectra are complicated by chemical disorder, lattice anharmonicity, and orientational averaging inherent to powder measurements. Guided by the symmetry-predicted mode classifications, we identify how the tetragonal-to-rhombohedral structural transformation is reflected in the evolution of the Raman spectra and how local symmetry breaking manifests as deviations from the average crystallographic description. This analysis further motivates a general strategy for interpreting disorder-broadened powder Raman spectra, in which the symmetry-predicted mode count provides a reference against which additional spectral features can be assessed. More broadly, the same workflow extends to other complex functional materials in which crystal symmetry governs both ferroic behavior and lattice dynamics.
	
	\section{Symmetry Workflow}
	\label{sec:workflow}
	
	As summarized in the overview of Fig.~\ref{fig:symmetry_workflow}, the analysis presents a systematic application of established crystallographic methods, coordinating the macroscopic and microscopic consequences of a single set of point-group operations. By doing so, we establish a mathematically consistent baseline that links macroscopic tensors to microscopic vibrations. For each crystallographic phase, the analysis begins with the space group and its associated point group, from which the complete set of symmetry operations is obtained. An orthogonal laboratory coordinate system is defined with respect to the crystallographic axes, enabling the construction of matrix representations of the symmetry operations. These matrices constitute the common mathematical foundation of the analysis.
	
	The central principle underlying both branches of the workflow is Neumann's principle, which states that every measurable physical property of a crystal must possess at least the symmetry of its point group. For a tensor of rank \(r\), \(T_{i_1\dots i_r}\), a symmetry operation represented by the direction-cosine matrix \(a_{ij}\) transforms the tensor according to
	
	\begin{equation}
		T'_{i_1 \dots i_r}
		=
		a_{i_1 j_1}\cdots a_{i_r j_r}
		T_{j_1 \dots j_r},
		\label{eq:tensor_transformation}
	\end{equation}
	
	\noindent
	where repeated indices imply summation. The invariance condition required by Neumann's principle,
	
	\begin{equation}
		T' = T,
	\end{equation}
	
	\noindent
	must be satisfied for every symmetry operation of the crystal point group. Within the macroscopic branch, general tensorial representations of physical properties such as pyroelectricity, dielectric susceptibility, and piezoelectricity are first constructed. The symmetry operations of the corresponding point group are then applied sequentially through Neumann's principle [Eq.~\eqref{eq:tensor_transformation}] to determine the nonzero independent tensor components permitted by the crystal symmetry.
	
	The microscopic branch applies the same symmetry framework to lattice vibrations. Atomic positions, Wyckoff-site symmetries, and the corresponding factor group are first identified. Because first-order Raman scattering probes only zone-center phonons (\(\mathbf{q}=0\)), translational symmetry reduces to the corresponding crystallographic point group. Using the associated point-group character tables, the total vibrational representation is then decomposed into irreducible representations through factor-group analysis (Bhagavantam--Venkatarayudu method) and correlation methods \cite{BornHuang1954,HayesLoudon1978,YuCardona2010}, as discussed in detail in the following sections. The optical vibrational representation is finally obtained by removing the acoustic contribution,
	
	\begin{equation}
		\Gamma_{\mathrm{opt}}
		=
		\Gamma_{\mathrm{vib}}
		-
		\Gamma_{\mathrm{acoustic}},
		\label{eq:optical_modes}
	\end{equation}
	
	\noindent
	thereby yielding the symmetry classification of the optical phonon modes. The basis functions listed in the character tables further determine the Raman- and infrared-activity selection rules. The final step of the microscopic branch is to determine the Raman tensors associated with the irreducible representations. Since Raman tensors are second-rank tensors, application of Neumann's principle similarly determines their symmetry-allowed forms. The character tables, basis functions, and correlation relations required for this analysis are available in standard group-theoretical references and crystallographic databases \cite{hamermesh1989group,cotton1991group,kroumova2003,Dresselhaus2008}. For completeness, the derivations relevant to the PZT phases considered in this work are presented in the Supplementary Information.
	
	Both branches ultimately yield experimentally measurable quantities associated with a given crystallographic phase. Changes in crystal symmetry modify both the allowed tensor forms and the vibrational spectra, providing complementary signatures of structural phase transitions. The analysis of these quantities forms the basis for the phase-specific investigations presented in the following sections.
	
	\section{Structural Phases of PZT}
	
\begin{figure*}[t]
		\centering
		\includegraphics[width=\linewidth]{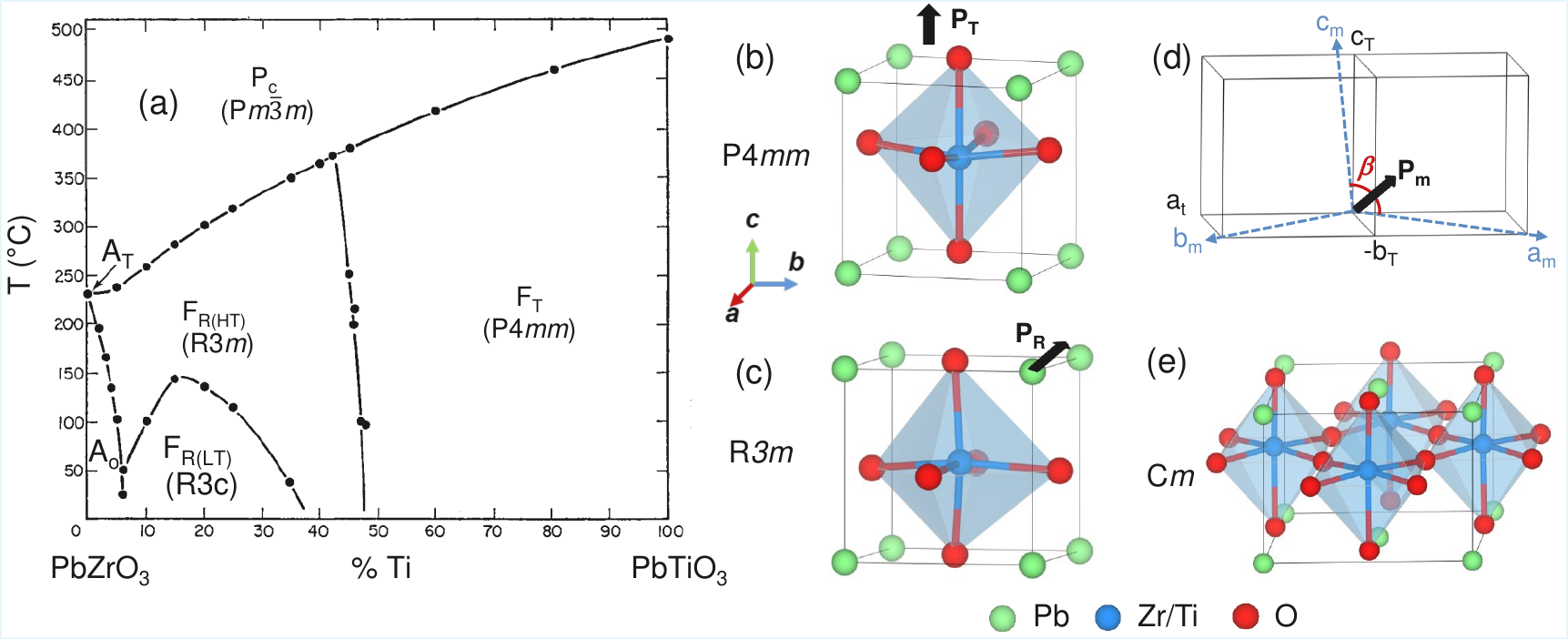}
		\caption{
			(a) Schematic phase diagram of PbZr$_x$Ti$_{1-x}$O$_3$ as a function of temperature and Ti concentration, expressed as $100(1 - x)\%$ Ti, adapted from \cite{Jaffe1971}: cubic (P$_C$, P$m\bar{3}m$), tetragonal (F$_T$, P$4mm$), and rhombohedral (F$_{R(\mathrm{HT})}$, R$3m$; F$_{R(\mathrm{LT})}$, R$3c$) phases. The MPB lies near $x \approx 0.52$, where a monoclinic (C$m$) phase is reported.
			(b) Tetragonal (P$4mm$) and (c) rhombohedral (R$3m$) unit cells in the pseudocubic setting, with polarization along [001] ($P_T$) and [111] ($P_R$), respectively.  
			(d) Orientation of the monoclinic axes relative to the tetragonal axes: $\mathbf{a}_m \parallel [\bar{1}\,\bar{1}\,0]$ and $\mathbf{b}_m \parallel [1\,\bar{1}\,0]$ ($\mathbf{a}_m \approx \mathbf{b}_m \approx \mathbf{a}_t / \sqrt{2}$), with $\mathbf{c}_m \approx \mathbf{c}_t$.
			(e) Schematic of the monoclinic unit cell (C$m$) of PZT near the MPB.
		}
		\label{fig:crystal_structure}
		
		\textbf{Alt text:} Phase diagram and schematic crystal structures of PbZr$_x$Ti$_{1-x}$O$_3$ showing tetragonal, rhombohedral, and monoclinic phases and their polarization directions.
		
	\end{figure*}

	Applying this general workflow to PZT first requires specifying its crystallographic input. This section therefore describes the structural phase diagram of PZT, shown schematically in Fig.~\ref{fig:crystal_structure}(a) \cite{Jaffe1971}, which encompasses multiple crystallographic symmetries that evolve sensitively with both composition and temperature \cite{Jaffe1971,Amin1981,AriGur1974,Buixaderas2015,Carl1971,Cordero2007,Corker1998,Deluca2011,Fernandes1995,Frantti2001,Hatch2002,Kornev2006,Noheda2001,nohedaMonoclinicPhasePZT2000,Noheda2000,Noheda1999,guo2000,li2016}. Here \(x\) denotes the Zr fraction in PbZr$_x$Ti$_{1-x}$O$_3$. The structural diversity of PZT originates from the competing tendencies of the Ti- and Zr-centered oxygen octahedra: Ti–O covalency stabilizes polar distortions and ferroelectric order, whereas Zr–O interactions favor antiferrodistortive and nonpolar configurations \cite{cordero2011,Yokota2009_Rhombohedral}. The balance between these effects produces symmetry-lowering transitions that strongly influence polarization orientation, lattice dynamics, and ferroic tensor properties.

	At high temperature, all PZT compositions adopt a centrosymmetric cubic perovskite
	structure (P$_C$, P$m\bar{3}m$, space group \#221), with $\text{Pb}^{2+}$ cations at the
	cube corners, $\text{Zr}^{4+}$/$\text{Ti}^{4+}$ cations at the body center, and
	$\text{O}^{2-}$ anions at the face centers \cite{Jaffe1971}. Upon cooling, different
	compositions develop distinct ferroelectric distortions. On the Ti-rich side
	($x \lesssim 0.52$), PZT transforms to a tetragonal phase (F$_T$, P$4mm$, space group
	\#99), in which off-centering
	of the cations and anions along [001] generates a spontaneous polarization parallel to the
	$c$-axis \cite{Noheda2001,nohedaMonoclinicPhasePZT2000,Noheda2000,Noheda1999}, as shown in
	Fig.~\ref{fig:crystal_structure}(b). In contrast, on the Zr-rich side
	($x \gtrsim 0.52$), PZT adopts a rhombohedral phase (F$_{R(\mathrm{HT})}$, R$3m$, space
	group \#160), where the polarization
	lies along the pseudocubic [111] direction
	\cite{Yokota2009_Rhombohedral,IUCrJ2018_LocalStructure}, as shown in
	Fig.~\ref{fig:crystal_structure}(c).
	
	Between these two regimes lies the MPB, a narrow compositional range near $x \approx 0.52$
	in which the tetragonal and rhombohedral phases become nearly degenerate in free energy
	\cite{Jaffe1971,Noheda2001,nohedaMonoclinicPhasePZT2000,Noheda2000,Noheda1999}. This
	near-degeneracy permits continuous rotation of the polarization between the [001] and [111]
	directions via intermediate monoclinic structures (C$m$, space group \#8)
	\cite{Noheda2001,nohedaMonoclinicPhasePZT2000,Noheda2000,Noheda1999}. The monoclinic
	C-centered unit cell is often described relative to the perovskite pseudocubic cell.
	Specifically, the monoclinic axes $\mathbf{a}_m$ and $\mathbf{b}_m$ lie along the
	tetragonal $\left[ \bar{1} \bar{1} 0 \right]$ and $\left[ 1 \bar{1} 0 \right]$ directions
	($\mathbf{a}_m \approx \mathbf{b}_m \approx \mathbf{a}_t / \sqrt{2}$), while $\mathbf{c}_m$
	is close to the $\left[ 0 0 1 \right]$ axis ($\mathbf{c}_m \approx \mathbf{c}_t$), as
	illustrated in Fig.~\ref{fig:crystal_structure}(d)--(e). The monoclinic cell has
	$\mathbf{b}_m$ as the unique axis, and the angle between $\mathbf{a}_m$ and $\mathbf{c}_m$
	is approximately $90.5^\circ$ at $20\,\text{K}$ \cite{Noheda1999}. The polarization rotation
	enabled by the monoclinic phase provides the microscopic basis for the enhanced
	piezoelectric and dielectric responses observed near the MPB, making PZT a benchmark
	material for high-performance piezoceramics
	\cite{Phelan2010_SingleCrystal,Burkovsky2012_Diffuse}. Detailed crystallographic data for
	these major phases are provided in Table S1 of the Supplementary Information.
	
	At lower temperatures, the high-temperature rhombohedral phase
	(F$_{R(\mathrm{HT})}$, R$3m$) undergoes a further transition to a low-temperature
	rhombohedral phase (F$_{R(\mathrm{LT})}$, R$3c$), in which cooperative oxygen octahedral
	tilting doubles the unit cell relative to the pseudocubic perovskite setting \cite{Yokota2009_Rhombohedral,Kocsis2014}. This transition introduces antiphase octahedral
	rotations described by the Glazer \(a^-a^-a^-\) tilt system. Because R$3m$ and R$3c$ share
	the same crystallographic point group, their macroscopic tensor forms are represented by
	the same rhombohedral symmetry treatment used here. However, the unit-cell doubling in
	R$3c$ folds additional zone-boundary modes to the zone center, and those translational
	symmetry effects are outside the scope of the present point-group framework.
	
	In addition to these major structural phases, other phases with different space group
	symmetries have also been reported \cite{li2016}. For the physical-property tensors and Raman analyses that
	follow, however, our discussion focuses on the cubic, tetragonal, rhombohedral
	R$3m$, and monoclinic C$m$ reference symmetries. For this
	purpose, we adopt a common orthogonal laboratory coordinate system $(x,y,z)$ for all
	phases. In the tetragonal phase, the $z$-axis is aligned with [001]; in the rhombohedral
	phase, it is aligned with [111], with the $x$-axis chosen in one of the vertical mirror
	planes. The monoclinic phase is described with its unique $b$-axis
	along $y$, corresponding to a mirror plane in the $xz$-plane \cite{cordero2011}. In this
	setting, the symmetry-operation matrices used for each phase, including the cubic
	generators, are provided in Section II of the Supplementary Information. This structural
	framework forms the basis for the symmetry analysis of the dielectric, piezoelectric,
	pyroelectric, and Raman tensor properties across the PZT phase diagram.

	\section{Macroscopic Physical Properties}
	\label{sec:tensors}
	
	This section applies the macroscopic branch of Fig.~\ref{fig:symmetry_workflow} to the four reference phases of PZT, treating the pyroelectric, dielectric, and piezoelectric tensors in order of increasing rank.
	
	\subsection{Pyroelectric Vector}
	
	Pyroelectricity is treated as a first-rank polar tensor constrained by Neumann's principle. The pyroelectric vector,
	\[
	\mathbf{p} = (p_x, p_y, p_z),
	\]
	must therefore remain invariant under all symmetry operations of the corresponding crystal point group, which directly determines the allowed nonzero components in each phase.
	
	In the rhombohedral ferroelectric phase (R3$m$), enforcing invariance ($\mathbf{p}' = \mathbf{p}$) under the threefold rotation about the [111] axis gives
	\[
	\begin{pmatrix}
		p_x' \\[1mm] p_y' \\[1mm] p_z'
	\end{pmatrix}
	=
	\begin{pmatrix}
		-\tfrac{1}{2} & -\tfrac{\sqrt{3}}{2} & 0 \\
		\tfrac{\sqrt{3}}{2} & -\tfrac{1}{2} & 0 \\
		0 & 0 & 1
	\end{pmatrix}
	\begin{pmatrix}
		p_x \\[1mm] p_y \\[1mm] p_z
	\end{pmatrix},
	\]
	resulting in $p_x = p_y = 0$, leaving only $p_z \neq 0$. Applying the remaining symmetry operations of the R3$m$ point group imposes no additional constraints on the nonzero component. The pyroelectric vector therefore lies along the [111] direction, coincident with the spontaneous polarization axis.
	
	In the tetragonal phase (P4$mm$), the fourfold rotation about [001] imposes the same condition, $p_x = p_y = 0$, leaving $p_z$ as the only nonzero component along the $c$-axis. By contrast, in the monoclinic phase (C$m$), which contains a mirror plane in the $xz$ plane, $p_y$ must vanish, but $p_x$ and $p_z$ are symmetry-allowed. The pyroelectric vector can therefore rotate freely within the $xz$ plane, bridging the tetragonal [001] and rhombohedral [111] polar directions. In the cubic paraelectric phase (P$m\bar{3}m$), inversion symmetry forbids any spontaneous polarization, yielding $\mathbf{p} = 0$.
	
	Reported pyroelectric coefficients in PZT films and ceramics are commonly of order \(10^{-4}\,\mathrm{C\,m^{-2}\,K^{-1}}\). In the units used here, representative values for device-relevant PZT compositions are approximately
	\(0.010\)--\(0.045~\mu\mathrm{C\,cm^{-2}\,^{\circ}C^{-1}}\), with many MPB-related reports falling near \(0.010\)--\(0.030~\mu\mathrm{C\,cm^{-2}\,^{\circ}C^{-1}}\)
	\cite{kesimPyroelectricResponseLead2013,Eklund2025_Pyroelectric_PZT}. The coefficient is highly sensitive to sample morphology and thermal history. Variations in factors such as grain size, domain structure, and crystallographic texture strongly influence the orientation and mobility of polar regions, thereby causing significant sample-to-sample variation in the measured coefficients \cite{Kamel2007_PyroVsCond}.
	
	\subsection{Dielectric Permittivity}
	
	The dielectric response is described by the second-rank tensor $\boldsymbol{\varepsilon}$, which relates the electric displacement $\mathbf{D}$ and electric field $\mathbf{E}$ as
	\begin{equation}
		D_i = \varepsilon_{ij} E_j.
		\label{eq:dielectric_relation}
	\end{equation}
	
	Under a symmetry operation represented by the transformation matrix $\mathbf{a}$, Neumann's principle requires
	\begin{equation}
		\boldsymbol{\varepsilon}' = \mathbf{a}\, \boldsymbol{\varepsilon}\, \mathbf{a}^{\mathrm{T}} = \boldsymbol{\varepsilon},
		\label{eq:invariance}
	\end{equation}
	which constrains the number of independent tensor components allowed in each crystallographic phase \cite{Nye1985}.
	
	\subsubsection*{Rhombohedral Phase (R3m)}
	
	In the rhombohedral phase, invariance under a threefold rotation about the [111] axis [Eq.~\eqref{eq:invariance}] reduces the tensor to the uniaxial form
	\begin{equation}
		\boldsymbol{\varepsilon} =
		\begin{pmatrix}
			\varepsilon_{\perp} & 0 & 0 \\[1mm]
			0 & \varepsilon_{\perp} & 0 \\[1mm]
			0 & 0 & \varepsilon_{\parallel}
		\end{pmatrix}.
	\end{equation}
	Here, $\varepsilon_{\parallel}$ and $\varepsilon_{\perp}$ are defined relative to the spontaneous polarization direction, reflecting dielectric anisotropy along and perpendicular to [111].
	
	\subsubsection*{Tetragonal Phase (P4mm)}
	
	Applying fourfold rotation symmetry about [001] and the associated mirror planes yields the same uniaxial form,
	\begin{equation}
		\boldsymbol{\varepsilon} =
		\begin{pmatrix}
			\varepsilon_{\perp} & 0 & 0 \\[1mm]
			0 & \varepsilon_{\perp} & 0 \\[1mm]
			0 & 0 & \varepsilon_{\parallel}
		\end{pmatrix}.
	\end{equation}
	The tetragonal phase is therefore also uniaxial, with anisotropy governed by the polar axis along $c$ \cite{Karapuzha2016_DielPZT}.
	
	\subsubsection*{Monoclinic Phase (Cm)}
	
	In the monoclinic phase, the reduced symmetry (single mirror plane in the $xz$ plane) allows coupling between $x$ and $z$ components, giving
	\begin{equation}
		\boldsymbol{\varepsilon} =
		\begin{pmatrix}
			\varepsilon_{xx} & 0 & \varepsilon_{xz} \\[1mm]
			0 & \varepsilon_{yy} & 0 \\[1mm]
			\varepsilon_{xz} & 0 & \varepsilon_{zz}
		\end{pmatrix}.
	\end{equation}
	The off-diagonal term $\varepsilon_{xz}$ reflects dielectric coupling between orthogonal axes and enables polarization rotation within the $xz$ plane, which is central to the enhanced dielectric response near the MPB \cite{guo2000,Roleder2022_MonoDiel,Zhang2014_MissingBoundary}.
	
	\subsubsection*{Cubic Phase (Pm$\bar{3}$m)}
	
	In the cubic paraelectric phase, full rotational symmetry enforces isotropy,
	\begin{equation}
		\boldsymbol{\varepsilon} =
		\varepsilon
		\begin{pmatrix}
			1 & 0 & 0 \\[1mm]
			0 & 1 & 0 \\[1mm]
			0 & 0 & 1
		\end{pmatrix}.
	\end{equation}
	
	As in the pyroelectric case, the measured dielectric response of PZT is strongly influenced by composition, microstructure, and processing conditions \cite{Karapuzha2016_DielPZT,Roleder2022_MonoDiel,Zhu2025_ThicknessDielectric}.
	
	\subsection{Piezoelectric Coupling}
	
	Piezoelectricity is described by a third-rank tensor $d_{ijk}$ that couples mechanical stress to electric polarization. Under the direct effect, stress generates polarization according to
	\begin{equation}
		P_i = d_{ijk} \, \sigma_{jk},
		\label{eq:piezotensor}
	\end{equation}
	where $\sigma_{jk} = \sigma_{kj}$ implies that $d_{ijk}$ is symmetric in its last two indices.
	
	In Voigt notation, this becomes
	\begin{equation}
		P_i = d_{im} \, \sigma_m,
		\label{eq:piezo_voigt}
	\end{equation}
	with $m=1\ldots6$ representing the independent stress components. Using this convention, the piezoelectric tensor is expressed as a $3\times6$ matrix,
	\begin{equation}
		\mathbf{d} =
		\begin{pmatrix}
			d_{11} & d_{12} & d_{13} & d_{14} & d_{15} & d_{16} \\
			d_{21} & d_{22} & d_{23} & d_{24} & d_{25} & d_{26} \\
			d_{31} & d_{32} & d_{33} & d_{34} & d_{35} & d_{36}
		\end{pmatrix}.
	\end{equation}
	
	As in Sec.~\ref{sec:workflow}, Neumann’s principle requires invariance under the symmetry operations of the crystal,
	\begin{equation}
		d'_{ijk} = a_{il} a_{jm} a_{kn} d_{lmn} = d_{ijk},
		\label{eq:piezo_invariance}
	\end{equation}
	which determines the symmetry-allowed tensor components in each phase.
	
	\subsubsection*{Rhombohedral Phase (R3m)}
	
	With the $x$ axis chosen in one of the mirror planes (the $xz$ plane), invariance under the threefold rotation along [111] and mirror operations [Eq.~\eqref{eq:piezo_invariance}] yields, in Voigt ordering $(xx,yy,zz,yz,xz,xy)$,
	\begin{equation}
		\mathbf{d} =
		\begin{pmatrix}
			-d_{22} & d_{22} & 0 & 0 & d_{15} & 0 \\
			0 & 0 & 0 & d_{15} & 0 & 2d_{22} \\
			d_{31} & d_{31} & d_{33} & 0 & 0 & 0
		\end{pmatrix}.
	\end{equation}
	Four independent coefficients remain: $d_{15}$, $d_{22}$, $d_{31}$, and $d_{33}$.
	
	\subsubsection*{Tetragonal Phase (P4mm)}
	
	The tetragonal symmetry yields
	\begin{equation}
		\mathbf{d} =
		\begin{pmatrix}
			0 & 0 & 0 & 0 & d_{15} & 0 \\
			0 & 0 & 0 & d_{15} & 0 & 0 \\
			d_{31} & d_{31} & d_{33} & 0 & 0 & 0
		\end{pmatrix},
	\end{equation}
	with three independent coefficients: $d_{15}$, $d_{31}$, and $d_{33}$.
	
	\subsubsection*{Monoclinic Phase (Cm)}
	
	For monoclinic symmetry with a mirror plane in $xz$, components containing an odd number of $y$ indices vanish:
	\begin{equation}
		\mathbf{d} =
		\begin{pmatrix}
			d_{11} & d_{12} & d_{13} & 0 & d_{15} & 0 \\
			0 & 0 & 0 & d_{24} & 0 & d_{26} \\
			d_{31} & d_{32} & d_{33} & 0 & d_{35} & 0
		\end{pmatrix}.
	\end{equation}
	Ten independent coefficients remain, reflecting enhanced electromechanical coupling under reduced symmetry.
	
	\subsubsection*{Cubic Phase (Pm$\bar{3}$m)}
	
	In the centrosymmetric cubic phase,
	\begin{equation}
		d_{ijk} = 0 \quad \forall i,j,k,
	\end{equation}
	and no piezoelectric response exists.
	
	Across all three properties, the number of symmetry-allowed independent components increases as the symmetry is lowered from cubic to monoclinic, as summarized in Table~\ref{tab:tensor_summary}. This systematic opening of response channels, together with the polarization rotation enabled by the monoclinic phase, underlies the enhanced pyroelectric, dielectric, and electromechanical responses observed near the MPB \cite{guo2000,fu2000,Shi2023_PhaseDiagram}.
	
	\begin{table}[t]
		\centering
		\begin{threeparttable}
			\caption{Number of symmetry-allowed independent components of the pyroelectric ($p_i$), dielectric ($\varepsilon_{ij}$), and piezoelectric ($d_{ij}$) tensors in the major PZT phases. The polarization direction indicates the symmetry-allowed polar axis in each phase.}
			\label{tab:tensor_summary}
			
			\renewcommand{\arraystretch}{1.2}
			\begin{tabular}{l c c c c c}
				\hline
				Phase & Point group & Polar axis & $p_i$ & $\varepsilon_{ij}$ & $d_{ij}$ \\
				\hline
				Cubic (P$m\bar{3}m$) & $m\bar{3}m$ & none & 0 & 1 & 0 \\
				Tetragonal (P$4mm$) & $4mm$ & $[001]$ & 1 & 2 & 3 \\
				Rhombohedral (R$3m$) & $3m$ & $[111]$ & 1 & 2 & 4 \\
				Monoclinic (C$m$) & $m$ & in mirror ($xz$) plane & 2 & 4 & 10 \\
				\hline
			\end{tabular}
			
			\begin{tablenotes}
				\footnotesize
				\item $d_{ij}$ denotes the number of independent piezoelectric coefficients in Voigt notation.
			\end{tablenotes}
			
		\end{threeparttable}
	\end{table}

	\section{Microscopic Vibrational Properties}
	\label{sec:vibrational}
	
	We now apply the microscopic branch of the workflow to classify the lattice vibrations of each PZT phase. While the group-theoretical classification of zone-center phonons is a well-established formalism, presenting it in parallel with the macroscopic tensor analysis provides an essential reference baseline for interpreting the experimental Raman spectra. This analysis classifies the lattice vibrations within the symmetry framework introduced in Sec.~\ref{sec:workflow}.
	
	Lattice vibrations (phonons) arise from collective atomic displacements around equilibrium positions and play a central role in dielectric response, electromechanical coupling, and structural stability in perovskite oxides \cite{LinesGlass1977,Nakamura1966_SoftModel,kamba2021}. In ferroelectrics, low-energy optical modes are particularly significant, as their instability is directly linked to the emergence of spontaneous polarization.
	
	Within the same symmetry framework, each normal mode transforms according to an irreducible representation (irrep) of the crystal point group. The corresponding character tables specify how atomic displacement patterns transform under symmetry operations and determine the degeneracy and optical activity of each mode \cite{Dresselhaus2008,HayesLoudon1978,cotton1991group,YuCardona2010}. These tables therefore provide a direct connection between crystal symmetry and vibrational spectra.
	
	To obtain the vibrational spectrum of a given phase, the full reducible vibrational representation is constructed and decomposed into irreps using standard group-theoretical procedures, such as the Bhagavantam--Venkatarayudu method \cite{Bhagavantam1964_GroupTheory,Adams1970_FactorGroup} or correlation analysis \cite{hornig1948,fateley1971}. When symmetry is lowered across a phase transition, this decomposition changes accordingly, leading to splitting of degeneracies and activation of additional Raman- or infrared-active modes.
	
	Raman spectroscopy provides an experimental probe of these symmetry-determined vibrations. Because Raman activity requires modulation of the electronic polarizability, changes in the Raman spectra directly reflect symmetry breaking, local distortions, or phase coexistence.
	
	In perovskite oxides such as PZT, structural transitions involve subtle displacements and tilts of oxygen octahedra and cations. Symmetry analysis therefore links these structural changes to the mode splitting and intensity redistribution observed in Raman spectra across the cubic, tetragonal, rhombohedral, and monoclinic phases. Detailed character tables, irreducible representations, and optical activities for the PZT phases are given in Section III of the Supplementary Information (Tables S2--S5), with additional derivations in Section IV (Tables S6--S14). In the following, we apply this framework explicitly to the rhombohedral phase before extending it to the remaining phases.

	\subsection{\label{sec:bv_method} Bhagavantam--Venkatarayudu (BV) Method}
	
	To ensure a self-contained and rigorous treatment, we employ two complementary, standard group-theoretical approaches: the Bhagavantam–Venkatarayudu (BV) factor-group method and the Wyckoff site-symmetry correlation method. The BV method, also known as factor-group analysis, is an established approach for determining the symmetries of the lattice vibrations of a crystal \cite{Bhagavantam1964_GroupTheory,Adams1970_FactorGroup,MITRA1961,deangelis1972}. In this method, one studies how each symmetry operation in the factor group affects each type of atom in the unit cell. The method requires knowledge of the crystal structure, the number of atoms in the unit cell, and the symmetry operations of the space group.
	
	A \emph{factor group} is obtained by separating the translational symmetry of the space
	group from its rotational and other symmetry operations. In practice, the infinite set of
	lattice translations is treated as the identity operation. This leaves only the rotational,
	reflection, inversion, screw, and glide symmetries that relate atoms within a single unit
	cell. The resulting group is isomorphic to one of the 32 crystallographic point groups and
	can be used to describe the unit cell symmetry in vibrational and spectroscopic analyses.
	
	For each symmetry operation \(R\) in the factor group, let \(\omega_R\) denote the number
	of atoms that remain unchanged by \(R\). Each invariant atom contributes to the trace of
	the vibrational displacement matrix, denoted as \(\chi_p(R)\). For Cartesian coordinates,
	this trace is given by \cite{rousseau1981}
	\begin{equation}
		\chi_p(R) = \omega_R (\pm 1 + 2\cos\theta),
	\end{equation}
	where \(\theta\) is the rotation angle of \(R\). The plus sign applies to proper rotations
	(without reflection), while the minus sign applies to improper rotations (involving
	inversion, reflection, or rotation-reflection).
	
	Once \(\chi_p(R)\) is calculated for all symmetry operations, the number of vibrational
	modes transforming as a given irreducible representation \(\Gamma_\gamma\) is obtained
	using
	\begin{equation}
		n^{(\gamma)} = \frac{1}{g} \sum_i g_i \chi_i^{(\gamma)} \chi_{p_i}(R),
	\end{equation}
	where \(g\) is the order of the group, \(g_i\) is the number of operations in the \(i\)-th
	class, \(\chi_i^{(\gamma)}\) is the character of that class for \(\Gamma_\gamma\), and
	\(\chi_{p_i}(R)\) is defined earlier. This approach systematically determines the
	symmetry of all vibrational modes in the crystal. By combining it with the character table,
	one can identify which modes are Raman- or infrared-active.

	\begin{table}
		\centering
		\caption{Application of the Bhagavantam--Venkatarayudu (BV) method to the rhombohedral phase of PZT (point group \(3m\)).}
		\label{tab:BV_rhombohedral}
		\renewcommand{\arraystretch}{1.2}
		\begin{tabular}{c c c c c}
			\hline
			Class & \(E\) & \(2C_3\) & \(3\sigma_v\) & \(n^{(\gamma)}\) \\
			\hline
			\(\omega_R\) & 5 & 2 & 3 & \\
			\(\theta\) & \(0^\circ\) & \(120^\circ\) & \(0^\circ\) & \\
			\(\chi_p\) & 15 & 0 & 3 & \\
			\hline
			\(\chi_i^{(A_1)}\) & 1 & 1 & 1 & \(\frac{1}{6}(15 + 0 + 9) = 4\) \\
			\(\chi_i^{(A_2)}\) & 1 & 1 & -1 & \(\frac{1}{6}(15 + 0 - 9) = 1\) \\
			\(\chi_i^{(E)}\) & 2 & -1 & 0 & \(\frac{1}{6}(30 + 0 + 0) = 5\) \\
			\hline
		\end{tabular}
	\end{table}
	Table~\ref{tab:BV_rhombohedral} applies these expressions to the
	rhombohedral structure (point group \(3m\)) of PZT. Under the identity operation, all five
	atoms remain unchanged, so \(\omega_R = 5\) and \(\theta = 0\). The complete analysis yields the total
	vibrational representation
	\[
	\Gamma_\mathrm{vib} = 4A_1 + A_2 + 5E,
	\]
	corresponding to 15 vibrational degrees of freedom: five singly degenerate modes and five
	doubly degenerate modes for the five-atom pseudocubic unit cell. After subtracting the
	three acoustic modes (\(A_1 + E\)) following Eq.~\eqref{eq:optical_modes}, the optical modes are distributed as
	\[
	\Gamma_\mathrm{opt} = 3A_1 + A_2 + 4E.
	\]
	Among these, the \(A_1\) and \(E\) modes are both Raman and infrared active, while the
	\(A_2\) mode is silent in first-order Raman and infrared spectra.

	\subsection{\label{sec:corr_method} Correlation Method}
	
	Complementing this, the correlation method offers a highly intuitive alternative by mapping local site symmetries directly to the crystal's factor group \cite{hornig1948,winston1949,fateley1971,tuschel2015}. While the BV method requires explicit evaluation of each symmetry operation (which can be laborious for complex structures), the correlation method uses the Wyckoff site symmetries of atoms in the unit cell to derive the distribution of vibrational modes.
	
	In this method, atomic vibrations are regarded as small displacements of atoms from their
	equilibrium positions, resolved along the crystallographic axes (\(x, y, z\)). Each
	displacement transforms according to an irreducible representation of the site group
	corresponding to that Wyckoff position. These site-group irreps are then correlated with
	the irreps of the factor group using standard correlation tables \cite{rousseau1981,fateley1971,CorrelationTables,kroumova2003}. The total vibrational representation of the crystal is obtained as
	\begin{equation}
		\Gamma_{\text{crystal}} = \Gamma_{\text{atom}_1} + \Gamma_{\text{atom}_2} + \cdots,
	\end{equation}
	where \(\Gamma_{\text{atom}}\) is the representation of atomic displacements at a given
	Wyckoff site.
	
	The number of translational modes transforming as site-group irrep \(\gamma\) is denoted
	\(t^\gamma\), and the number of rotational modes is \(R^\gamma\). If \(n\) is the number of
	equivalent atoms occupying a given Wyckoff position, then the degrees of vibrational
	freedom for that site are
	\begin{equation}
		f_T^\gamma = n \cdot t^\gamma,
	\end{equation}
	and
	\begin{equation}
		f_R^\gamma = n \cdot R^\gamma.
	\end{equation}
	The correlation table gives how the translational species of a site-symmetry irrep
	\(\gamma\) decompose into factor-group irreps \(\zeta\). If \(a_\zeta\) is the number of
	times a factor-group irrep \(\zeta\) occurs for a given Wyckoff site containing \(n\)
	equivalent atoms, the dimensionality constraint is
	\begin{equation}
		3n = \sum_\zeta a_\zeta C_\zeta,
		\label{eq:corr_dimension}
	\end{equation}
	where \(C_\zeta\) is the degeneracy of the factor-group species \(\zeta\). The values of
	\(C_\zeta\) correspond to the degeneracies of the symmetry species: 1 for \(A\) or \(B\), 2
	for doubly degenerate \(E\), and 3 for triply degenerate \(T\) species. Summing the
	contributions from all occupied Wyckoff sites gives the total number \(a_\zeta\) of lattice
	vibrations transforming as each factor-group irrep \(\zeta\).
	
	The total irreducible representation of the crystal, \(\Gamma_{\text{crystal}}\), can thus
	be constructed as
	\begin{equation}
		\Gamma_{\text{crystal}} = \sum_\zeta a_\zeta \cdot \zeta.
		\label{eq:gamma_crystal}
	\end{equation}

	In the rhombohedral structure, the unit cell contains one formula unit and one lattice
	point. The Wyckoff sites are \(1a\) for the Pb and Zr/Ti atoms \((n=1)\) and \(3b\) for the
	oxygen atoms \((n=3)\), as summarized in Table S1 of the Supplementary Information. Because
	the correlation tables use Schoenflies notation to denote both the site-symmetry groups and
	the factor group, we follow the same notation here.
	
	\begin{itemize}
		\item For the Pb and Zr/Ti atoms at the \(1a\) site, the correlation table for \(C_{3v} \rightarrow C_{3v}^{5}\) is used, as shown in Table \ref{tab:PbZrTi_1a}.
		\item For the O atoms at the \(3b\) site, the correlation table for \(C_{s} \rightarrow C_{3v}^{5}\) is used, as shown in Table \ref{tab:O_3b}.
	\end{itemize}

	\begin{table*}
		\centering
		\caption{Correlation between the site group \(C_{3v}\) and the factor group \(C_{3v}^5\) for the Pb and Zr/Ti atoms (\(1a\) sites, \(n = 1\)) in rhombohedral PZT.}
		\label{tab:PbZrTi_1a}
		\includegraphics[width=0.7\linewidth]{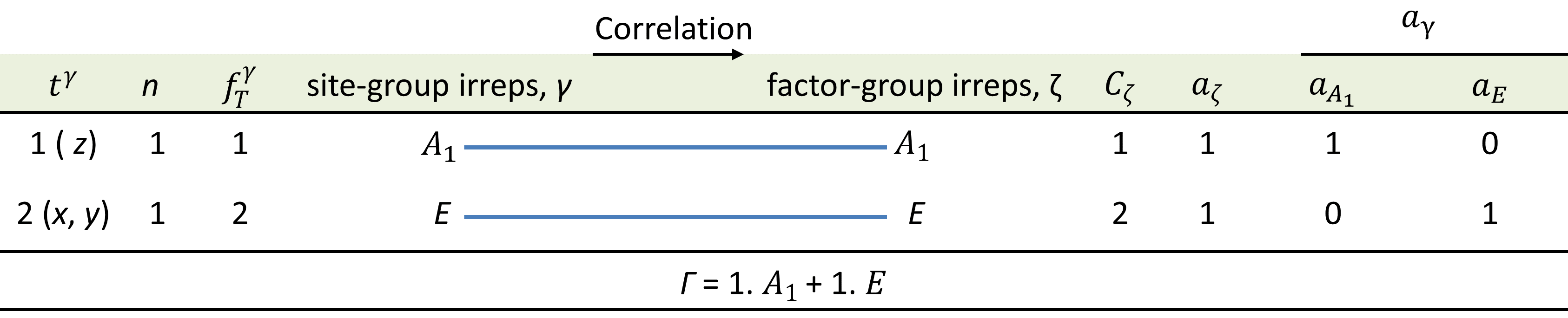}
	\end{table*}

	\begin{table*}
		\centering
	\caption{Correlation between the site group \(C_s\) and the factor group \(C_{3v}^5\) for the O atoms (\(3b\) site, \(n = 3\)) in rhombohedral PZT.}
		\label{tab:O_3b}
		\includegraphics[width=0.7\linewidth]{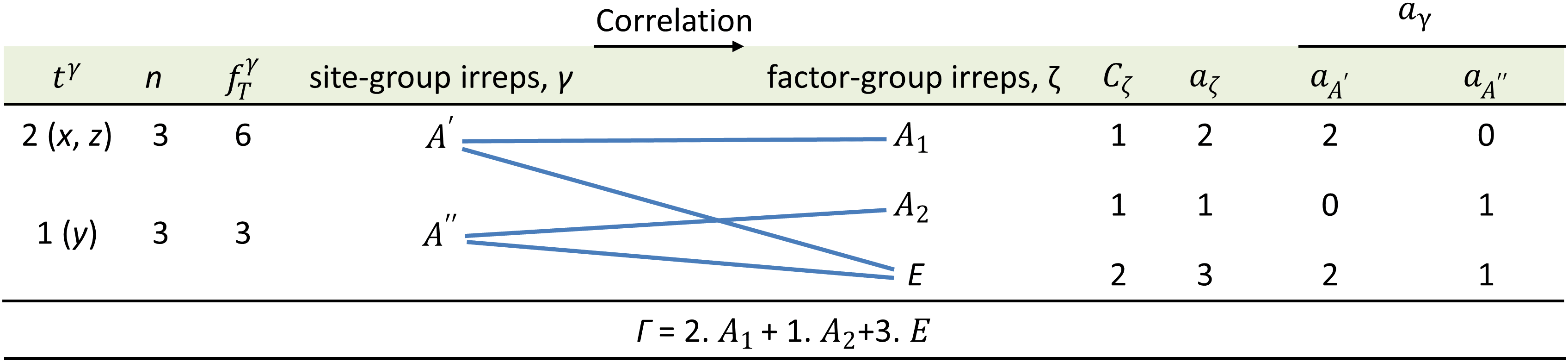}
	\end{table*}

	The first step is to identify the site-group irreps associated with translational
	displacements and then correlate them with the factor-group irreps using
	Eqs.~\eqref{eq:corr_dimension} and \eqref{eq:gamma_crystal}. For each \(1a\) site, the translational representation is \(A_1 + E\). Since
	both Pb and Zr/Ti occupy \(1a\) sites, their combined contribution is
	\begin{align}
		\Gamma_{\mathrm{Pb}}+\Gamma_{\mathrm{Zr/Ti}}
		= 2(A_1 + E).
	\end{align}
	
	For the oxygen atoms at the \(3b\) site, the local \(C_s\) species correlate with the
	\(C_{3v}\) factor group as \(A' \rightarrow A_1 + E\) and \(A'' \rightarrow A_2 + E\). The
	three oxygen atoms therefore contribute
	\begin{align}
		\Gamma_{\mathrm{O}}
		= 2(A_1 + E) + (A_2 + E)
		= 2A_1 + A_2 + 3E.
	\end{align}
	
	Therefore, the total vibrational representation of the crystal is
	\begin{equation}
		\Gamma_{\text{crystal}} = 2(A_1 + E) + (2A_1 + A_2 + 3E) = 4A_1 + A_2 + 5E.
	\end{equation}
	After subtracting three acoustic modes \((A_1 + E)\) following Eq.~\eqref{eq:optical_modes}, the optical modes are
	\begin{equation}
		\Gamma_{\text{optical}} = 3A_1 + A_2 + 4E,
	\end{equation}
	which matches precisely the result obtained from the BV method. The agreement between the two independent methods provides a useful internal check, and the same procedure can be applied to any crystal once its space group and Wyckoff occupations are known.

	\section{Raman Tensors}
	\label{sec:raman_tensors}
	
	With the vibrational symmetries established, the Raman tensors are constructed as the final step of the microscopic branch of the workflow. This step follows directly from the symmetry-determined irreducible representations by treating the Raman tensor as a second-rank polar tensor constrained by Neumann's principle.
	
	In Raman scattering, the energy difference between the incident and scattered photons corresponds to a phonon frequency, with phonon creation and annihilation producing the Stokes and anti-Stokes lines, respectively, as illustrated in Fig.~\ref{fig:Raman_process}(a). The scattering intensity arises from the modulation of the electronic polarizability tensor $\boldsymbol{\alpha}$ by lattice vibrations. The induced dipole moment is
	\begin{equation}
		p_i = \sum_j \alpha_{ij} E_j,
	\end{equation}
	and expanding the polarizability with respect to a phonon coordinate $q_k$ gives
	\begin{equation}
		\alpha_{ij} = \alpha_{ij}^0 + \left( \frac{\partial \alpha_{ij}}{\partial q_k} \right)_0 q_k + \cdots.
	\end{equation}
	
	The Raman tensor of mode $k$ is defined as
	\begin{equation}
		R_{ij}^{(k)} = \left( \frac{\partial \alpha_{ij}}{\partial q_k} \right)_0,
	\end{equation}
	which directly connects lattice displacements to optical polarizability changes~\cite{loudon2001,pimenta2021,TubinoPiseri1975}. The Raman intensity for a given experimental configuration is then
	\begin{equation}
		I \propto \left| \mathbf{e}_s \cdot \mathbf{R} \cdot \mathbf{e}_i \right|^2.
	\end{equation}
	
	To determine the allowed form of $\mathbf{R}$ for each mode symmetry, we enforce Neumann's principle in tensor form. The Raman tensor is treated as a general $3\times3$ matrix constrained to transform as the corresponding irrep under all symmetry operations of the crystal point group,
	
	\begin{equation}
		\mathbf{a}\, \mathbf{R}\, \mathbf{a}^\mathrm{T} = \mathbf{R},
		\label{eq:raman_neumann}
	\end{equation}
	
	This procedure eliminates symmetry-forbidden components and yields the basis set of Raman tensors associated with each vibrational mode~\cite{alfaro2011,Benshalom2023_PhononPhononRaman,keresztury2006}.
	
	In ferroelectric perovskites such as PZT, symmetry lowering from the cubic to tetragonal, rhombohedral, and monoclinic phases reduces degeneracies and activates additional Raman modes, providing a direct spectroscopic signature of structural transitions.
	
	\subsection*{Rhombohedral phase}
	
	For the rhombohedral phase, the symmetry is $3m$, with the $z$ axis chosen along the threefold polar direction and the $xz$ plane as a mirror plane. The Raman tensors are obtained by applying the symmetry operations of this point group to a general second-rank tensor and enforcing invariance [Eq.~\eqref{eq:raman_neumann}] within each irreducible representation.
	
	For the totally symmetric $A_1$ mode, this procedure yields a diagonal tensor,
	\[
	R(A_1) =
	\begin{pmatrix}
		a & 0 & 0 \\
		0 & a & 0 \\
		0 & 0 & b
	\end{pmatrix}.
	\]
	
	For the doubly degenerate $E$ modes, as derived in detail in the Supplementary Information, the symmetry constraints generate a two-dimensional tensor basis,
	\[
	R(E_x) =
	\begin{pmatrix}
		c & 0 & d \\
		0 & -c & 0 \\
		d & 0 & 0
	\end{pmatrix},
	\quad
	R(E_y) =
	\begin{pmatrix}
		0 & -c & 0 \\
		-c & 0 & d \\
		0 & d & 0
	\end{pmatrix}.
	\]
	
	These forms follow directly from the reduction of a general tensor under the $3m$ symmetry constraints and define the polarization selection rules for the rhombohedral phase.
	
	\subsection*{Tetragonal phase}
	
	For the tetragonal phase ($P4mm$), the same procedure is applied using fourfold rotational symmetry and vertical mirror planes. The resulting optical representation is $\Gamma_\mathrm{opt} = 3A_1 + B_1 + 4E$, where all modes are Raman active.
	
	The symmetry constraints yield
	
	\[
	R(A_1) =
	\begin{pmatrix}
		a & 0 & 0 \\
		0 & a & 0 \\
		0 & 0 & b
	\end{pmatrix}, \quad
	R(B_1) =
	\begin{pmatrix}
		c & 0 & 0 \\
		0 & -c & 0 \\
		0 & 0 & 0
	\end{pmatrix},
	\]
	
	\[
	R(E_x) =
	\begin{pmatrix}
		0 & 0 & e \\
		0 & 0 & 0 \\
		e & 0 & 0
	\end{pmatrix},
	\quad
	R(E_y) =
	\begin{pmatrix}
		0 & 0 & 0 \\
		0 & 0 & f \\
		0 & f & 0
	\end{pmatrix}.
	\]
	
	The selection rules follow directly from the allowed tensor components: $A_1$ and $B_1$ modes appear in parallel-polarized geometries, while $E$ modes require polarization components involving the $z$ axis.
	
	\subsection*{Monoclinic phase}
	
	In the monoclinic phase, the symmetry is reduced to a single mirror plane, leaving fewer constraints on the tensor form. The same symmetry projection procedure yields $\Gamma_\mathrm{opt} = 7A' + 5A''$, with both irreps Raman active.
	
	The Raman tensors are
	
	\[
	R(A') =
	\begin{pmatrix}
		a & 0 & d \\
		0 & b & 0 \\
		d & 0 & c
	\end{pmatrix},
	\quad
	R(A'') =
	\begin{pmatrix}
		0 & e & 0 \\
		e & 0 & f \\
		0 & f & 0
	\end{pmatrix}.
	\]
	
	The reduced symmetry increases the number of independent tensor components, allowing additional coupling between polarization directions.
	
	\subsection*{Cubic phase}
	
	In the cubic phase ($Pm\bar{3}m$), the same construction applied to the high-symmetry point group yields $\Gamma_\mathrm{opt} = 3T_{1u} + T_{2u}$. None of these modes transform according to Raman-active irreducible representations.
	
	Consequently, all Raman tensor components are symmetry-forbidden at the zone center, and first-order Raman scattering is absent. Any experimentally observed Raman signal in this phase must therefore arise from symmetry-breaking effects such as local disorder, polar fluctuations, or nanoscale distortions.
	
	\begin{table*}[t]
		\centering
		\caption{Raman-active irreps, their tensor representations, and polarization configurations, given in Porto notation (defined in Sec.~\ref{sec:raman_exp}), for the major PZT phases. \(X\), \(Y\), and \(Z\) are the phase-specific Cartesian axes; for the rhombohedral phase, \(Z\parallel[111]\) with \(X\) in a vertical mirror plane. The monoclinic phase is treated in the one-formula-unit pseudocubic setting. The Raman-active vibrational representation is listed below each phase name. Full mode decompositions are given in the text and the Supplementary Information.}
		\label{tab:RamanPZT}
		\small
		\renewcommand{\arraystretch}{1.3}
		\setlength{\tabcolsep}{5pt}
		\begin{tabular}{@{}lcllc@{}}
			\toprule
			Phase & Irrep(s) & Raman Tensor $\mathbf{R}$ & Symmetry Type & Typical Configurations \\
			\midrule
			
			\shortstack[l]{Rhombohedral (R$3m$)\\[2pt] {\footnotesize Raman: $3A_1+4E$}} & $A_1$ &
			$\begin{pmatrix} a & 0 & 0 \\ 0 & a & 0 \\ 0 & 0 & b \end{pmatrix}$ &
			symmetric &
			$Z(XX)\bar{Z}$, $Z(YY)\bar{Z}$, $X(ZZ)\bar{X}$ \\
			& $E$ &
			$\begin{pmatrix} c & 0 & d \\ 0 & -c & 0 \\ d & 0 & 0 \end{pmatrix}$,
			$\begin{pmatrix} 0 & -c & 0 \\ -c & 0 & d \\ 0 & d & 0 \end{pmatrix}$ &
			degenerate &
			$Z(XX)\bar{Z}$, $Z(XY)\bar{Z}$, $Y(XZ)\bar{Y}$, $X(YZ)\bar{X}$ \\
			\midrule
			
			\shortstack[l]{Tetragonal (P$4mm$)\\[2pt] {\footnotesize Raman: $3A_1+B_1+4E$}} & $A_1$ &
			$\begin{pmatrix} a & 0 & 0 \\ 0 & a & 0 \\ 0 & 0 & b \end{pmatrix}$ &
			symmetric &
			$Z(XX)\bar{Z}$, $Z(YY)\bar{Z}$, $X(ZZ)\bar{X}$ \\
			& $B_1$ &
			$\begin{pmatrix} c & 0 & 0 \\ 0 & -c & 0 \\ 0 & 0 & 0 \end{pmatrix}$ &
			singly degenerate &
			$Z(XX)\bar{Z}$, $Z(YY)\bar{Z}$, $Z(X'Y')\bar{Z}$ \\
			& $E$ &
			$\begin{pmatrix} 0 & 0 & e \\ 0 & 0 & 0 \\ e & 0 & 0 \end{pmatrix}$,
			$\begin{pmatrix} 0 & 0 & 0 \\ 0 & 0 & f \\ 0 & f & 0 \end{pmatrix}$ &
			degenerate &
			\(Y(XZ)\bar{Y}\), \(X(YZ)\bar{X}\) \\
			\midrule
			
			\shortstack[l]{Monoclinic (C$m$)\\[2pt] {\footnotesize Raman: $7A'+5A''$}} & $A^{\prime}$ &
			$\begin{pmatrix} a & 0 & d \\ 0 & b & 0 \\ d & 0 & c \end{pmatrix}$ &
			mirror-even &
			$Z(XX)\bar{Z}$, $Z(YY)\bar{Z}$, $X(ZZ)\bar{X}$, $Y(XZ)\bar{Y}$ \\
			& $A^{\prime\prime}$ &
			$\begin{pmatrix} 0 & e & 0 \\ e & 0 & f \\ 0 & f & 0 \end{pmatrix}$ &
			mirror-odd &
			$Z(XY)\bar{Z}$, $X(YZ)\bar{X}$ \\
			\midrule
			
			\shortstack[l]{Cubic (P\(m\bar{3}m\))\\[2pt] {\footnotesize Raman: none}} & \(T_{1u}\), \(T_{2u}\) &
			none &
			Raman inactive &
			first-order forbidden \\
			\bottomrule
		\end{tabular}
	\end{table*}

	Table~\ref{tab:RamanPZT} summarizes the Raman-active irreps, their tensor forms, and
	representative polarization geometries for each major phase of PZT. In summary, the cubic
	phase lacks first-order Raman modes, while the rhombohedral, tetragonal, and monoclinic
	phases exhibit progressively richer Raman spectra reflecting the symmetry lowering.
	Experimentally, these symmetry-dependent tensor forms enable phase identification and reveal
	structural transitions and polarization rotation near the morphotropic
	boundary.
	
	\begin{figure}[t]
		\centering
		\includegraphics[width=1\linewidth]{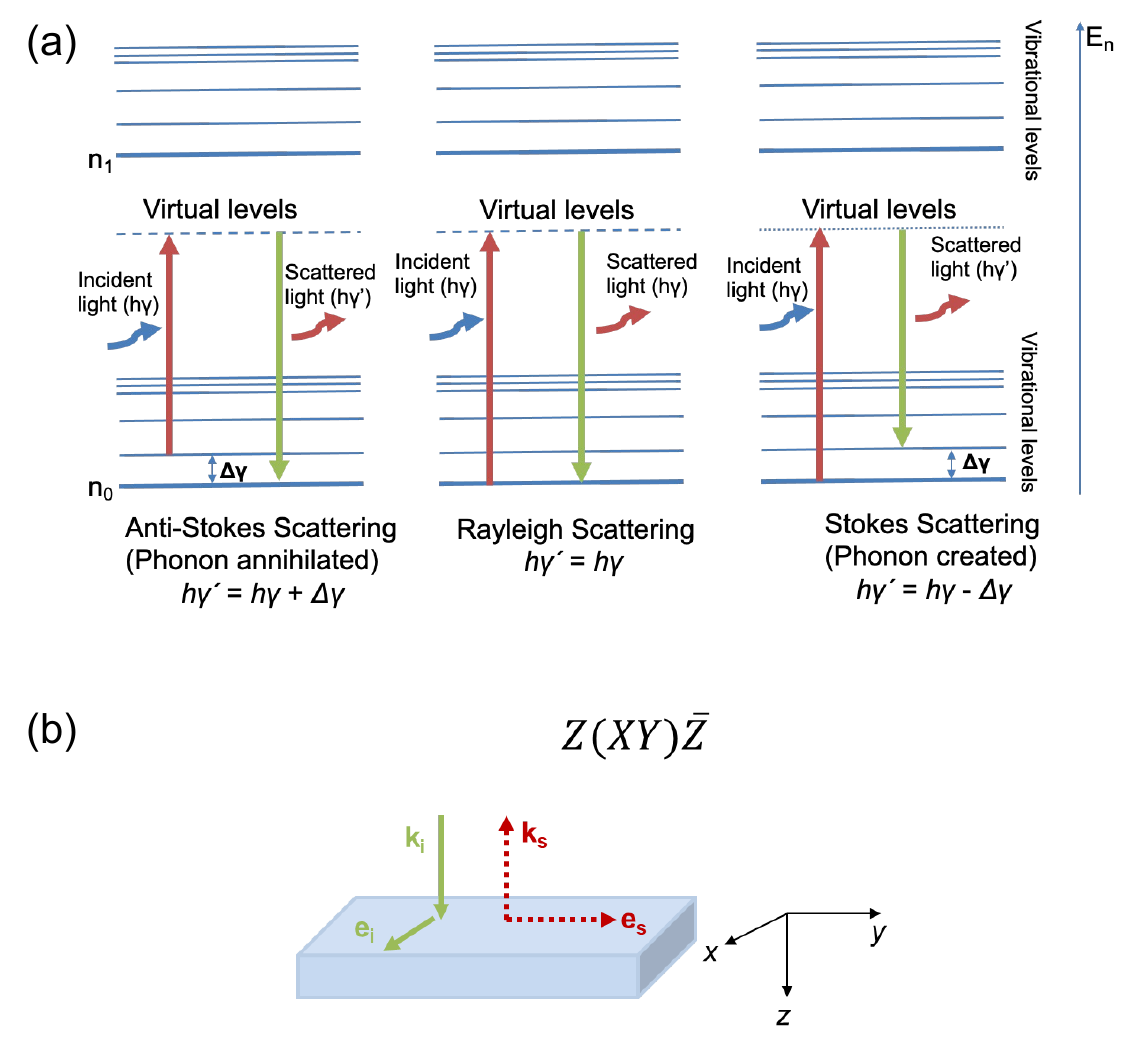}
		\caption{%
			(a) Raman (Stokes and anti-Stokes) and Rayleigh scattering processes in a crystal, illustrating phonon creation and annihilation.
			(b) Backscattering geometry for the crossed-polarization configuration $Z(XY)\bar{Z}$; $\mathbf{k}$ and $\mathbf{e}$ denote the propagation directions and polarization vectors of the incident ($i$) and scattered ($s$) light.
		}
		\label{fig:Raman_process}
		\textbf{Alt text:} Schematic of Raman scattering processes and backscattering measurement geometry showing crossed-polarization configuration.
	\end{figure}

	\section{Polarization Selection Rules}
	\label{sec:raman_exp}
	
	While the crystal symmetry fundamentally determines the Raman-active phonon modes, their
	experimental observation is strongly influenced by both the sample geometry and the
	polarization configuration \cite{yi2004,deluca2007,deluca2006,lee2002}. Careful
	optimization of these parameters is therefore required to selectively probe phonon modes with
	specific symmetries, thereby enabling clearer interpretation of phase-dependent spectral
	features.
	
	In Raman experiments, the polarization states of the incident and scattered light are
	commonly described using the Porto notation,
	\[
	\mathbf{k}_i \, (\mathbf{e}_i \, \mathbf{e}_s) \, \mathbf{k}_s,
	\]
	where $\mathbf{k}_i$ and $\mathbf{k}_s$ denote the propagation directions of the incident
	and scattered light, respectively, and $\mathbf{e}_i$ and $\mathbf{e}_s$ represent their
	corresponding polarization directions. For example, $Z(XX)\bar{Z}$ corresponds to a
	parallel polarization configuration, while $Z(XY)\bar{Z}$ represents crossed polarization.
	
	In a typical back-scattering configuration such as $Z(XX)\bar{Z}$, both the incident and
	scattered beams propagate along the $Z$-axis (with the scattered beam reflected along
	$-Z$), and their polarization vectors are aligned along the $X$-axis:
	\[
	\mathbf{e}_i = (1,0,0), \quad \mathbf{e}_s = (1,0,0),
	\]
	giving
	\[
	\mathbf{e}_s^{\mathrm{T}} \mathbf{R} \mathbf{e}_i = 
	(1,0,0)
	\begin{pmatrix}
		R_{xx} & R_{xy} & R_{xz} \\
		R_{yx} & R_{yy} & R_{yz} \\
		R_{zx} & R_{zy} & R_{zz}
	\end{pmatrix}
	\begin{pmatrix} 1 \\ 0 \\ 0 \end{pmatrix} = R_{xx}.
	\]
	Hence, the parallel polarization directly probes the diagonal tensor elements. In contrast, for a crossed configuration such as $Z(XY)\bar{Z}$, as shown in Fig.~\ref{fig:Raman_process}(b), where
	$\mathbf{e}_i = (1,0,0)$ and $\mathbf{e}_s = (0,1,0)$,
	\[
	\mathbf{e}_s^{\mathrm{T}} \mathbf{R} \mathbf{e}_i = R_{yx},
	\]
	which involves the off-diagonal components of the tensor.
	
	These geometries therefore select modes by symmetry: parallel configurations emphasize modes with nonzero diagonal Raman-tensor elements, such as $A_1$, $B_1$, and $A'$, whereas crossed configurations reveal modes with off-diagonal elements, such as $E$ and $A''$, depending on the phase. Table~\ref{tab:RamanPZT} summarizes representative polarization orientations in which these modes can be observed for the major PZT phases. In practice, the table can be read as an experimental design guide: choosing a scattering geometry isolates specific mode symmetries and allows phase assignments to be tested directly. Direct experimental verification of these symmetry-based selection rules requires single crystals, poled ceramics, or epitaxial thin films aligned along high-symmetry directions (e.g., $Z \parallel [001]$ or $[111]$).

	As a concrete example, the tetragonal $B_1$ mode offers a clean symmetry marker of the tetragonal-to-rhombohedral transition. In a crossed backscattering geometry such as $Z(X'Y')\bar{Z}$, $B_1$ is allowed while the $E$ modes, whose Raman tensors contain only $xz$ and $yz$ components, are extinguished. Thereby, isolating $B_1$ from the $E$ mode with which it overlaps on the powder data. Because the rhombohedral $3m$ point group contains no $B_1$ irrep, tracking the $B_1$ intensity with composition on an oriented single crystal would directly follow the loss of tetragonal symmetry across the morphotropic phase boundary.
	
	In polycrystalline or powder samples, however, random grain orientations average out the polarization dependence of Raman scattering. The measured spectra then contain all symmetry-allowed modes of the phase but cannot by themselves provide unique mode-symmetry assignments, which instead require polarization-resolved measurements on oriented samples or \emph{ab initio} calculations. Modes with diagonal Raman-tensor elements typically dominate the intensity, whereas those with only off-diagonal elements appear weaker and orientation-averaged.
	
	When Raman-active phonons are simultaneously infrared-active, they are called as polar modes like $A_{1}$ and $E$. In these modes, coupling between the dipole-induced electric field and ionic motion gives rise to longitudinal--transverse optical (LO--TO) splitting. The LO vibration corresponds to polarization parallel to the phonon propagation vector, while the TO vibration corresponds to polarization perpendicular to it. Because longitudinal atomic vibrations generate a macroscopic electric field that opposes and stiffens the motion, an additional restoring force is introduced, shifting the LO phonon frequency higher relative to the TO mode~\cite{durman1987}. The magnitude of this splitting is described by the Lyddane--Sachs--Teller (LST) relation \cite{lyddane1941}. For a material such as PZT that supports multiple infrared-active polar modes, the LST relation generalizes to a product over all polar mode pairs:
		\begin{equation}
			\label{eq:LST}
			\frac{\varepsilon(0)}{\varepsilon(\infty)} =
			\prod_{j} \left( \frac{\omega_{\mathrm{LO},j}}{\omega_{\mathrm{TO},j}} \right)^{2},
		\end{equation}
		where the product runs over all zone-center infrared-active modes $j$, each characterized by a transverse optical frequency $\omega_{\mathrm{TO},j}$ and a corresponding longitudinal optical frequency $\omega_{\mathrm{LO},j}$. Each polar mode therefore contributes independently to the ratio of the static to high-frequency dielectric constant. Thus, the analysis of LO-TO splitting provides a quantitative connection between lattice dynamics and the dielectric behavior of functional materials.

	\begin{figure}[t]
		\centering
		\includegraphics[width=0.95\linewidth]{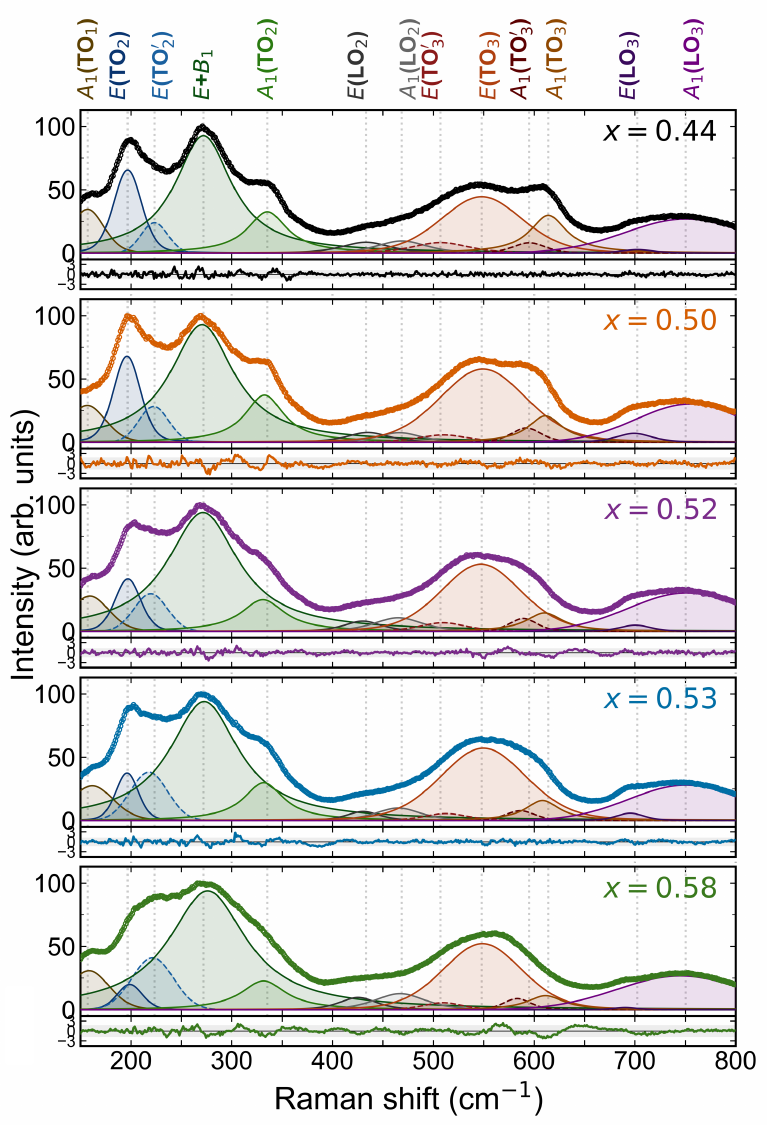}
		\caption{%
			Room-temperature powder Raman spectra of PbZr$_x$Ti$_{1-x}$O$_3$ across the morphotropic phase boundary (MPB) for compositions $x = 0.44$–$0.58$, reproduced from Ref.~\cite{maity2023}. Vertical dashed lines mark the principal phonon components identified from the symmetry-constrained decomposition.
		}
		\label{fig:PowderSpectraData}
		\textbf{Alt text:} Room-temperature powder Raman spectra of PbZr$_x$Ti$_{1-x}$O$_3$ showing composition-dependent changes in phonon modes across the morphotropic phase boundary.
	\end{figure}

	\begin{figure}[t]
		\centering
		\includegraphics[width=0.7\linewidth]{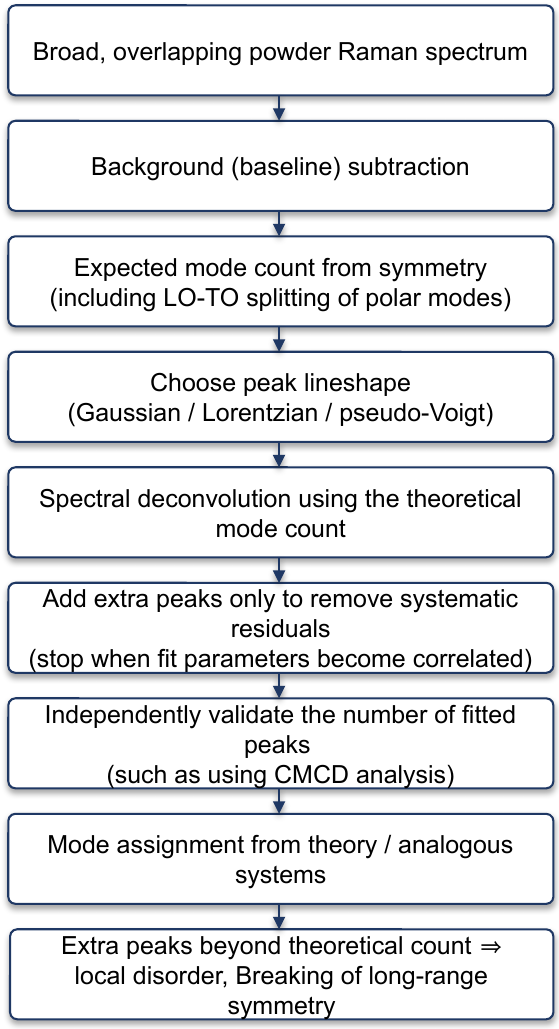}
		\caption{General workflow for analyzing disorder-broadened powder Raman spectra. The symmetry-predicted mode count provides a minimal, constrained peak decomposition that is cross-checked independently (e.g., CMCD). Extra peaks beyond the ideal count signal local disorder that breaks ideal long-range symmetry.}
		\label{fig:raman_workflow}
		\textbf{Alt text:} Flowchart showing the data analysis workflow for disorder-broadened powder Raman spectra, illustrating the path from experimental measurement and symmetry-constrained deconvolution to the identification of local disorder.
	\end{figure}

	\section{Powder Raman Data of PZT}
	\label{sec:pzt_results}
	
	\subsection{Fitting Strategy}
	
	Figure~\ref{fig:PowderSpectraData} shows room-temperature powder Raman spectra of PZT across the MPB for compositions $x = 0.44$–$0.58$, measured using 532~nm excitation. These spectra were previously reported in Ref.~\cite{maity2023} and are reanalyzed here within the group-theoretical framework developed in the preceding sections. X-ray diffraction on the same samples reveals a structural evolution from tetragonal ($x = 0.44$) to rhombohedral ($x = 0.58$) symmetry through a coexistence region ($x = 0.50, 0.52, 0.53$) centered near the MPB~\cite{maity2023}, consistent with its sensitivity to long-range crystallographic order. In contrast, Raman spectroscopy is particularly sensitive to local structural distortions, short-range order, and disorder-induced symmetry breaking that may not be visible in diffraction. The present analysis focuses on how experimental Raman spectra reflect this local structural response across the MPB, in comparison with the phase coexistence picture obtained from diffraction.
	
	The observed spectra can be broadly divided into three frequency regions~\cite{Burns1970,Freire1988}. The low-frequency region
	($\sim50$--$150$~cm$^{-1}$) corresponds to the Last mode, associated primarily with vibrations of the
	A-site Pb cation against the relatively rigid oxygen octahedral framework. The intermediate-frequency
	region ($\sim200$--$400$~cm$^{-1}$) is dominated by Slater-type modes, arising from vibrations of the
	B-site cation (Ti/Zr) against the oxygen octahedra. The high-frequency region ($\sim400$--$800$~cm$^{-1}$)
	is associated with Axe-type modes, corresponding mainly to oxygen-octahedral bending and internal
	oxygen vibrations.
	
	While the preceding symmetry analysis establishes the idealized crystallographic framework for Raman activity in ordered ferroelectric perovskites, Raman scattering from PZT near the MPB is largely influenced by additional complexities arising from local structural disorder and compositional heterogeneity. Random Zr/Ti occupation of the B-site, together with correlated Pb off-centering, produces a distribution of local environments that deviate from the ideal tetragonal and rhombohedral symmetries~\cite{Kakegawa1977,Kakegawa1995,Mishra1996}. Extended factor-group analysis for off-centered Pb displacements indicates that such local symmetry breaking increases the number of locally Raman-active vibrational configurations beyond those expected for an ideal ordered crystal~\cite{Buixaderas2008,Frantti2013}. At the same time, local variations in bond lengths and force constants introduce substantial inhomogeneous broadening, in addition to intrinsic phonon lifetime effects. Consequently, the measured spectra consist of broad, overlapping envelopes rather than well-resolved normal modes. This makes direct mode assignment from the raw spectra unfeasible, requiring a physically constrained decomposition. The general strategy for such data analysis is summarized in Fig.~\ref{fig:raman_workflow}. The tetragonal $x = 0.44$ sample exhibits the sharpest spectral features and is used as the reference.
	
	To account for both homogeneous lifetime broadening (Lorentzian contribution) and inhomogeneous disorder-induced broadening (Gaussian contribution)~\cite{Frantti1999,Frantti2013}, the Raman lineshapes were modeled using pseudo-Voigt functions,
	\begin{equation}
		I_j(\omega)=A_j\left[\eta_jL_j(\omega)+(1-\eta_j)G_j(\omega)\right],
	\end{equation}
	where
	\begin{align}
		L_j(\omega) &=
		\frac{(\Gamma_j/2)^2}
		{(\omega-\omega_{0j})^2+(\Gamma_j/2)^2}, \\
		G_j(\omega) &=
		\exp\!\left[
		-\frac{(\omega-\omega_{0j})^2}{2\sigma_j^2}
		\right], \qquad
		\sigma_j=
		\frac{\Gamma_j}{2\sqrt{2\ln2}}.
	\end{align}
	Here, $\omega_{0j}$ denotes the mode center frequency, $\Gamma_j$ the full width at half maximum (FWHM), $A_j$ the mode amplitude, and $\eta_j$ the Lorentzian--Gaussian mixing parameter. A linear baseline $B(\omega)=B_0+k\omega$ was subtracted prior to fitting. The mixing parameter $\eta_j$ was determined from unconstrained fits to the reference $x=0.44$ composition and subsequently fixed across all compositions to ensure consistency of the decomposition in the strongly overlapping spectral regions. The resulting fits yield RMS residuals of 0.53--0.69\% without systematic structure across the measured spectral range.
	
	\subsection{Spectral Deconvolution}
	
	From group-theoretical considerations, the tetragonal ($\Gamma_{\text{optical}} = 3A_1 + B_1 + 4E$) and rhombohedral ($\Gamma_{\text{optical}} = 3A_1 + 4E$) phases of PZT have the same number of zone-center vibrational modes for the five-atom perovskite unit cell. These mode counts do not differentiate between the two phases, and therefore a tetragonal-to-rhombohedral transition is not expected to involve the creation or loss of Raman-active phonon branches. This provides the basis for the spectral decomposition scheme.
	
	For tetragonal PZT and PbTiO$_3$, previous studies~\cite{Burns1970,Burns1973,Freire1988,Buixaderas2015,Deluca2011,Buixaderas2011,Frantti1996,Rouquette2006} show that the ideal structure supports up to 13 Raman-active components due to LO--TO splitting. Out of these 13 modes, only approximately ten should be experimentally resolvable, owing to spectral overlap and low-frequency cutoffs of our data. However, our least-squares fitting shows that a 13-component decomposition is required to avoid structured residuals across all compositions, as shown in Fig.~\ref{fig:PowderSpectraData}. These additional components do not correspond to new irreducible representations but possibly arise from disorder-induced splitting of symmetry-allowed modes. Our final decomposition comprises one Last-band component
	[$A_1(\mathrm{TO}_1)$], four Slater-band components
	[$E(\mathrm{TO}_2)$, $E(\mathrm{TO}_2')$, $E{+}B_1$,
	$A_1(\mathrm{TO}_2)$], six Axe-band components
	[$E(\mathrm{LO}_2)$, $A_1(\mathrm{LO}_2)$,
	$E(\mathrm{TO}_3')$, $E(\mathrm{TO}_3)$,
	$A_1(\mathrm{TO}_3')$, $A_1(\mathrm{TO}_3)$], and two LO$_3$
	components [$E(\mathrm{LO}_3)$, $A_1(\mathrm{LO}_3)$]. The same decomposition basis remains adequate for all compositions up to $x = 0.58$ without requiring the introduction of additional peaks into the fitting process. 
	
	The concave-down curvature-maxima (CMCD) analysis~\cite{Buixaderas2015}, which assumes no specific lineshape model, is applied to the reference $x = 0.44$ composition as an independent check on the adopted decomposition basis, as shown in Figure S1 of the Supplementary Information. In this approach, hidden spectral components are identified from extrema in the
	second-derivative curvature spectrum satisfying the conditions $d_3 = 0$ and $d_2 < 0$, where $d_n = d^nI/d\omega^n$, corresponding to local concave-down maxima
	in the Raman intensity profile. The first derivative $d_1=dI/d\omega$ identifies the principal intensity extrema, while the second and third derivatives isolate weak or strongly overlapping
	contributions that are not directly visible in the raw spectrum. Prior to derivative analysis, the Raman spectra were smoothed using a Savitzky--Golay filter while preserving the underlying spectral features.
	
	The CMCD analysis identifies ten reliable curvature maxima below
	$\sim650$~cm$^{-1}$ at approximately 153, 164, 197, 222, 268, 330,
	431, 462, 515, and 546~cm$^{-1}$, together with additional features
	at 199, 337, 547, 606, and 614~cm$^{-1}$ located in regions of
	stronger spectral overlap. Above $\sim650$~cm$^{-1}$ the
	third-derivative spectrum becomes progressively noise dominated,
	preventing reliable curvature identification in the high-frequency
	LO$_3$ region. Nevertheless, two clear LO$_3$ intensity maxima remain
	directly observable in the measured spectrum near
	743~cm$^{-1}$. Overall, the CMCD results support, though do not uniquely determine, the presence of approximately 13 underlying spectral contributions and provide an independent consistency check on the adopted decomposition basis.
	
	\subsection{Mode Splitting and Local Disorder}
	
	A notable feature of the decomposition is the subpeak structure of selected Raman modes within both the Slater and Axe bands, present throughout the entire composition range and most prominent in the $E(\mathrm{TO}_2)$/$E(\mathrm{TO}_2')$, $E(\mathrm{TO}_3)$/$E(\mathrm{TO}_3')$, and $A_1(\mathrm{TO}_3)$/$A_1(\mathrm{TO}_3')$ mode pairs. The splitting of the transverse $E$-symmetry modes arises from local symmetry breaking, driven by the statistical distribution of B-site cations and correlated Pb off-centering \cite{Frantti1999,Frantti2013}. In ideal tetragonal symmetry the $E$ modes are doubly degenerate transverse vibrations, but local variations in the B-site environment lift this degeneracy and generate multiple locally distinct vibrational configurations. Frantti \textit{et al.} showed that such splitting reflects spatially varying local order in PZT and cannot be described within a simple homogeneous soft-mode picture~\cite{Frantti2013}. The observed $E(\mathrm{TO})$ doublets are therefore consistent with local symmetry breaking and configurational heterogeneity within the MPB region.
	
	The origin of the
	$A_1(\mathrm{TO}_3)$/$A_1(\mathrm{TO}_3')$
	splitting is less straightforward. Unlike the strongly anharmonic
	low-frequency $A_1(\mathrm{TO}_1)$ soft mode discussed extensively for
	PbTiO$_3$ and related
	perovskites~\cite{Burns1973,Frantti1999,Frantti1997},
	the present splitting occurs in a higher-frequency Axe-band
	vibration dominated primarily by oxygen-octahedral motions along the
	polar axis. The additional shoulder near
	$\sim595$~cm$^{-1}$ therefore likely reflects local variations in
	Ti--O bond environments and polar-axis force constants arising from
	the combined effects of chemical disorder, local strain, and
	anharmonic polar distortions. In this sense, the observed
	$A_1(\mathrm{TO}_3)$ substructure may be viewed as
	phenomenologically analogous to the anharmonic subpeak structure
	reported for the $A_1(\mathrm{TO}_1)$ soft mode. The persistence of these split or shoulder-like
	features across all compositions supports the interpretation that local structural
	heterogeneity remains an intrinsic characteristic of the MPB region
	rather than a localized anomaly associated with a single composition.
	
	\subsection{Mode Evolution}
	
	Figure S2 of the Supplementary Information shows the composition dependence of the fitted peak positions, linewidths, and amplitudes. The phonon evolution is continuous across the MPB, with all Raman branches varying smoothly with composition and without abrupt discontinuities, mode splitting into new branches, or emergence of additional symmetry-lowering modes. Within experimental resolution, the structural transformation is therefore reflected primarily in a redistribution of spectral weight rather than qualitative changes in the phonon spectrum, consistent with a gradual evolution between competing local vibrational environments.
	
	The phonon frequencies exhibit only moderate composition dependence. The polar $A_1(\mathrm{TO}_2)$ mode softens gradually from 335 to 331~cm$^{-1}$, while the $A_1(\mathrm{TO}_3')$ shoulder exhibits the clearest monotonic redshift, decreasing from 595 to 583~cm$^{-1}$. Both trends indicate progressive weakening of the tetragonal polar-axis distortion with increasing Zr content. The LO$_3$ modes also soften systematically, reflecting the increasing average B-site mass and modification of Ti--O force constants. In contrast, several $E$-symmetry modes remain nearly stationary, consistent with the weaker sensitivity of transverse vibrations to B-site substitution. None of the phonon branches exhibits anomalous softening or discontinuous behavior near the MPB.
	
	The linewidth evolution reveals a different aspect of the MPB physics. Several modes, particularly $A_1(\mathrm{TO}_1)$, $A_1(\mathrm{TO}_2)$, and $A_1(\mathrm{LO}_2)$, exhibit maximal broadening near $x\approx0.52$--$0.53$, indicating enhanced structural heterogeneity and increased fluctuations of the local polar environment within the coexistence region. In contrast, the disorder-sensitive $E(\mathrm{TO}_2')$ and $E{+}B_1$ modes broaden progressively across the entire composition range, suggesting that substitutional disorder continues to increase even outside the nominal MPB composition.
	
	The clearest spectroscopic signature of the structural crossover is observed in the amplitude evolution. The main $E(\mathrm{TO}_2)$ component decreases strongly in intensity from approximately 66 to 20 arb. units, while the accompanying $E(\mathrm{TO}_2')$ shoulder simultaneously increases from approximately 24 to 41 arb. units. Rather than indicating a new phonon branch, this behavior reflects a continuous transfer of spectral weight between transverse modes as rhombohedral character increases. This spectral-weight transfer is the Raman counterpart of the tetragonal-to-rhombohedral transformation observed in diffraction; the coexistence region appears here not as two resolvable sets of phase-specific peaks but as the simultaneous presence of overlapping tetragonal-like and rhombohedral-like contributions within the same broad bands. A similar reduction is observed for the polar $A_1(\mathrm{TO}_2)$ mode, whose intensity decreases steadily across the composition range, consistent with suppression of the tetragonal polar-axis distortion.
	
	Although the decomposition reveals systematic and physically meaningful compositional trends, it is important to distinguish between quantitatively robust and qualitatively constrained parameters. The Slater-band parameters remain comparatively well separated and are therefore quantitatively reliable, whereas the strongly overlapping Axe-band region (430--620~cm$^{-1}$) contains correlated fitting parameters because of severe spectral overlap. The Axe-band amplitudes and linewidths should therefore be interpreted primarily in a qualitative sense. Nevertheless, the continuous evolution of all fitted parameters and the absence of systematic residual structure indicate that the adopted decomposition captures the essential spectral evolution consistently across the full composition range.
	
	\subsection{Possible Monoclinic Signatures}
	
	As the Raman spectra evolve smoothly across the morphotropic phase boundary (MPB) at room temperature without abrupt symmetry-breaking signatures, an important question is whether any of the observed features nevertheless indicate monoclinic symmetry. From the symmetry analysis presented above, a long-range monoclinic phase would be identified not by peak broadening alone, but by composition-dependent splitting of parent higher-symmetry modes. In particular, low-temperature Raman studies have shown that the region near $\sim280~\mathrm{cm}^{-1}$ provides one of the clearest fingerprints of monoclinic symmetry: on the tetragonal side, the $B_1/E$ feature reorganizes into monoclinic components, whereas on the rhombohedral side a previously silent $A_2$-derived mode becomes Raman active and the $E$ branch splits. Likewise, the accompanying reduction in intensity of the rhombohedral mode near $\sim240~\mathrm{cm}^{-1}$ has been associated with the rhombohedral-to-monoclinic crossover~\cite{Lima2001,SouzaFilho2002,SouzaFilho2000}. Consequently, the expected experimental signature of a long-range monoclinic phase is a reproducible $A'/A''$ doublet or multiplet, particularly for the $E(\mathrm{TO}_2)$ and $E(\mathrm{TO}_3)$ modes, rather than a general increase in linewidth.
	
	No such $A'/A''$ doublets or newly activated monoclinic branches are resolved for any composition in the present room-temperature powder spectra. Instead, the evolution across the MPB is characterized by a gradual redistribution of spectral weight between tetragonal-like and rhombohedral-like vibrational contributions, while the observed substructure of the $E$ modes is more consistently attributed to local symmetry breaking arising from chemical disorder and configurational heterogeneity~\cite{Frantti2013}. These conclusions are fully consistent with the comprehensive Raman investigation of Buixaderas \textit{et al.}~\cite{Buixaderas2015}, who examined PZT over a substantially broader temperature range (10--600 K), covered a lower-frequency region down to 20~cm$^{-1}$, and investigated a wide composition range ($0.25 \leq x \leq 0.70$). They likewise found no spectroscopic evidence for a long-range monoclinic phase, as indicated by the absence of systematic splitting of the doubly degenerate $E$ modes into monoclinic $A'$ and $A''$ components. Despite employing different fitting functions, both studies use derivative-based curvature analysis to constrain the spectral decomposition, leading to the same physical interpretation that the additional spectral features arise from local structural heterogeneity rather than from the emergence of a distinct crystallographic phase.
	
	These observations, however, do not exclude the existence of monoclinic nanodomains, adaptive monoclinic correlations, or local cation displacements~\cite{Woodward2005,Schonau2007,IUCrJ2018_LocalStructure,Wang2007}. Rather, they indicate that any such distortions remain confined to short length scales, where they may contribute to diffraction signatures or be detected by local structural probes, but do not produce resolvable symmetry splitting in the disorder-averaged Raman response.
	
	Further insight into symmetry evolution near the MPB will require probes with enhanced symmetry selectivity or local structural sensitivity. Polarization-resolved Raman measurements on oriented samples could recover symmetry information lost through powder averaging, while low-temperature measurements would reduce thermal broadening and improve mode resolution. Pair distribution function analysis using neutron or synchrotron X-ray scattering would further provide direct access to short-range structural correlations. These investigations are beyond the scope of the present work but offer promising directions for future studies.

	\section{Conclusions}
	\label{sec:conclusions}
	
	In conclusion, we have demonstrated how the systematic application of established crystallographic methods provides a highly consistent reference baseline for analyzing functional materials. By applying this parallel workflow to PZT, we show how a single set of crystallographic point-group operations dictates both the allowed forms of macroscopic physical-property tensors and the symmetries of microscopic vibrational modes. Applied to PZT, the analysis shows an increase in the number of independent tensor components with symmetry lowering across the morphotropic phase boundary. Within this symmetry-based mode deconvolution, the room-temperature Raman spectra as a function of composition are consistently described using 13 phonon components. The tetragonal-to-rhombohedral transition manifests as a continuous redistribution of spectral intensity among these components without the emergence of new modes, and no clear signature of long-range monoclinic symmetry is observed. Subpeak structures of certain Raman modes are present in all compositions and therefore are likely associated with local symmetry breaking arising from disorder and anharmonicity. This work demonstrates that the systematic application of established symmetry methods provides a consistent reference baseline for interpreting structure-property relationships and local structural deviations in complex ferroic materials.

	\section*{Acknowledgments}
	
	The authors acknowledge the Department of Physics, Indian Institute of Technology Madras, India, for support of the Raman experiments on which this work is based.
	
	\section*{Conflict of Interest}
	The authors declare that they have no competing interests.
	
	\section*{Funding}
	No funding was received for this work.
	
	\bibliography{Manuscript}

\end{document}